\documentclass[runningheads]{llncs}
\usepackage{graphicx}
\usepackage{xcolor}
\usepackage[hidelinks]{hyperref}

\usepackage{booktabs}

\begin{document}
\title{On the Relevance of Blockchain Evaluations on Bare Metal}
%
\author{Andrei Lebedev\inst{1}\orcidID{0000-0003-1994-9163} \and
\\ Vincent Gramoli\inst{1,2}\orcidID{0000-0001-5632-8572}}
\authorrunning{A. Lebedev and V. Gramoli}
%
\institute{University of Sydney, Sydney, Australia\\
\email{andrei.lebedev@sydney.edu.au}\\
\and Redbelly Network, Sydney, Australia\\
\email{vincent.gramoli@sydney.edu.au}}
\maketitle              
\begin{abstract}
In this paper, we present the first bare metal comparison of modern blockchains, including Algorand, Avalanche, Diem, Ethereum, Quorum and Solana. This evaluation was conducted with the recent Diablo benchmark suite~\cite{gramoli_diablo_2023}, a framework to evaluate the performance of different blockchains on the same ground. By tuning  network delays in our controlled environment we were able to reproduce performance trends obtained in geo-distributed settings, hence demonstrating the relevance of bare metal evaluations to better understand blockchain performance. 

\end{abstract}

\section{Introduction}

In recent years, there has been a significant increase in the variety of available blockchain protocols, finding applications in various domains such as finance, supply chain management, and healthcare~\cite{wang_bft_2022}. These diverse use cases give rise to distinct system requirements, encompassing factors like participation types and transaction metrics, particularly when subjected to varying workloads. At the same time, these blockchain protocols are structured with multiple layers, including the membership selection layer, consensus layer, data layer, and execution layer~\cite{NYG19,fan_performance_2020}, each precisely tailored to address specific utilization scenarios.

In the process of selecting appropriate blockchain protocols, system developers are tasked with making well-informed decisions by considering the array of available options, each tailored to specific layers in line with their particular requirements. 

To facilitate this decision-making, benchmarking has emerged as a valuable tool to assess various systems~\cite{touloupou_systematic_2022}. Both protocol developers and researchers contribute to this evaluation process by reporting metrics such as transaction throughput, latency, and resource utilization. However, it is important to note that the evaluation environment itself can vary significantly across different experiments. These variations encompass a range of scenarios, from utilizing Internet of Things (IoT) devices to leveraging high-performance computing clusters~\cite{HSGX20}, which could be situated in a single datacenter or distributed across various remote locations.

Researchers evaluated the protocols in various experiments using bare metal clusters~\cite{saingre_bctmark_2020,nasrulin_gromit_2022,ren_bbsf_2023}, yet an area that remains relatively unexplored is the influence of network equipment on these setups. While the majority of the studies consider average latency~\cite{gramoli_diablo_2023}, blockchain experiments often overlook a notable aspect that is the tail latency~\cite{dean_tail_2013}, which can provide more comprehensive insights into performance. It is worth noting that some authors have taken steps to emulate a geo-distributed environment~\cite{RG23}, but a critical gap exists in terms of directly comparing results from these emulations to those obtained within a standard network, using identical workloads. Bridging this gap and conducting a side-by-side assessment of these two distinct environments under the same experimental conditions would yield valuable insights into the performance and viability of blockchain implementations across different network configurations.

\sloppy{In this paper, we present the first bare metal comparison of modern blockchains. To this end, we evaluate six blockchains,
Algorand~\cite{GHM17}, Avalanche~\cite{TR21}, Diem~\cite{BBC19}, Ethereum~\cite{Woo15}, Quorum~\cite{jpmorganchase-quorum} and Solana~\cite{Yak21}
with the recent Diablo~\cite{gramoli_diablo_2023} benchmark suite in a cluster. We make the following contributions:}
\begin{itemize}
\item The performance trends obtained on our cluster with artificial network delays are similar to the ones obtained on geo-distributed settings.
In the Testnet and Devnet configurations, Solana provides similar performance and this was observed on the cluster as well as in a geo-distributed virtualized environment of previous work. 
Diem provides higher throughput for Testnet than for Devnet as observed in both the cluster and the geo-distributed virtualized environment.
\item \sloppy{We show that switches in the LAN do not impact the performance of blockchain significantly. This is explained by the blockchain latencies being in the order of the second whereas the switches would impact services with latencies orders of magnitude smaller (of the order of the millisecond).}
\item We show that the average blockchain transaction latency is generally not representative of its tail latency. In particular, Algorand and Quorum would typically have a significantly larger tail latency than average latency under 1000\,TPS and 10,000\,TPS workloads. 
While Solana does not experience much difference in our experiments, the difference on Avalanche and Diem can be high as well.
\end{itemize}

This paper is organised as follows. Section~\ref{sec:env-comparison} describes the differences in local and cloud environments, and lists the benefits and the drawbacks of each setting. In Section~\ref{sec:evaluation}, we demonstrate the performance of the 6 blockchain protocols in the local testbed, focusing on the impact of network switches, number of blockchain nodes, and the network delay between them. Section~\ref{sec:eval-limit} looks at the different aspects of the evaluation which can be taken into account in order to increase the depth of understanding of the blockchain protocols. We discuss the related work in Section~\ref{sec:related-work} and conclude the paper in Section~\ref{sec:conclusion}.

\section{Analysis}\label{sec:env-comparison}

In this section, we look into the differences of cloud provided virtual machines and local testbeds on the example of Amazon Web Services (AWS) and the i8 chair testbed.

\subsection{Cloud Environments}

The cloud environments provide certain benefits for the blockchain protocol evaluation. One of the important points is scalability in terms of computing power. Amazon Web Services (AWS) provide a vast range of machine types, for example, from 2 vCPUs and 4\,GB RAM to 96 vCPUs and 192\,GB RAM. Protocols can be optimized for different hardware with multiprocessing capabilities, such as GPU and CPU with vector extensions or CPU with specific architecture. Such factors are taken into account by the providers, and machines with different hardware are also available.

While multiple machines of different types can be spawned in the same datacenter, cloud providers also typically have multiple datacenters across the globe. This brings us to the second benefit of AWS, which is geographical distribution. Currently, Amazon has AWS datacenters in more than ten regions. With this feature, we can create networks of hundreds of machines, which allows us to easily test the scalability aspect of the protocol in terms of the number of blockchain nodes.

The fact that the datacenters are distributed across the globe provides us with a network with realistic latency and bandwidth. Even though virtualization is present in the setup, the machines share the actual hardware and network links. As the datacenters are located on different continents, we are provided with the latencies limited by the physical properties of the connection and the actual distance and underlying network topology between the locations. As different services on the machines communicate with each other and are accessed by the users, the bandwidth of the links is being used. This allows taking another important aspect of real networks into account during the evaluation, which is background traffic. 

However, such an environment makes it hard to perform reproducible tests. The utilization of the network links between the datacenters changes throughout the day, as the services may be accessed more during the day and less at night. The usage is reflected in latency and bandwidth, with lower latency and higher bandwidth available at hours with reduced usage and higher round-trip time (RTT) and lower throughput being observed at peak usage hours.

\subsection{Local Testbeds}

In order to account for the variance in the network parameters, local testbeds can be used to perform the measurements. In this environment, the whole network can be exclusively used by the system under test. For example, with the iLab testbed, we have measured an average of 1.1 millisecond RTT using the same approach as with AWS. Given that the latency between the nodes does not change, we can introduce arbitrary delays to evaluate the tolerance of the blockchain protocol against the network delays.

Such property of the iLab testbed network as fixed latency between the nodes allows us to replicate the latencies of geographically distributed cloud networks at a particular point in time. With \texttt{tc-netem} tool, we can specify the added delay on a network interface of the machine used for running the experiments. Ideally, such a setup can be used to reduce the usage of cloud environments, produce similar results, and reduce the cost of the experiments.

In order for the deployment solution to be backend-independent, it should operate on a unified protocol, such as SSH. In this case, it will be possible to operate on any set of servers that are accessible with SSH on the host.

\section{Evaluation}\label{sec:evaluation}

In this section, we evaluate six blockchain protocols with Diablo on the local testbed environment.

Diablo~\cite{gramoli_diablo_2023} is a blockchain benchmarking framework that allows comparing different protocols on the same ground with realistic workloads based on real application traces. Diablo has master-worker architecture, where \textit{Primary} acts as an orchestrator and result aggregator, and \textit{Secondaries} produce the workload and collect results for individual transactions. Diablo is accompanied by a set of scripts called Minion~\cite{gramoli_vincent_2022_7707312} that allows to automate the experiments.

The rationale for choosing the blockchain protocols is that they represent various consensus algorithms and virtual machines with different properties. Avalanche and Algorand offer probabilistic consensus algorithms, Diem and Quorum use variants of deterministic Byzantine fault tolerant consensus algorithms, and Ethereum and Solana use eventually consistent consensus algorithms. From the virtual machine perspective, Avalanche, Quorum, and Ethereum use Ethereum Virtual Machine and Solidity programming language, which Solana also makes use of with the Solang compiler. Algorand features Transaction Execution Approval language and Algorand Virtual machine, and Diem provides smart contract capabilities with Move programming language and MoveVM.

\begin{figure}
\centering
  \includegraphics[keepaspectratio=true,width=\textwidth]{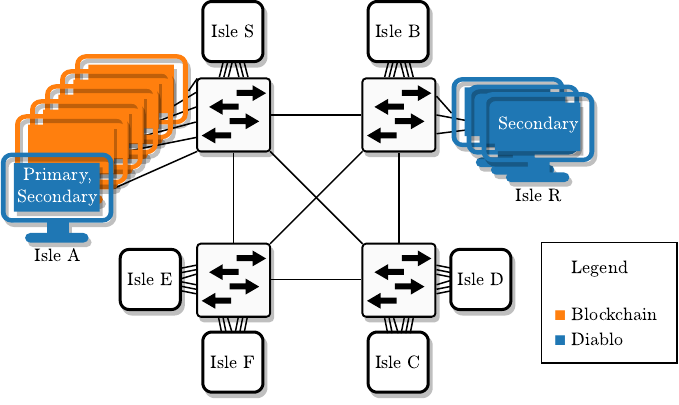}
  \caption{\label{fig:ilab}iLab topology}
\end{figure}

To run the experiments on the iLab~\cite{Pahl17} testbed, we use \texttt{eth-static} interface, which is a dedicated 10 gigabit network between all the testbed machines. As shown in Figure~\ref{fig:ilab}, the network consists of 7 groups of machines called isles (named A, B, C, D, E, F, S) of 6 machines each, plus an isle of 3 machines (named R), giving 8 isles and 45 machines in total. Every two isles (A and S, B and R, C and D, E and F) are connected to a switch, and there are overall 4 switches, and all of them are connected to each other. We interface Minion with Plain Orchestration Service~\cite{GallScho21} to allocate and deploy the machines.

The machines have the following hardware:
\begin{itemize}
	\item Intel Core i7-8700 CPU @ 3.20GHz (6/12 cores/threads)
	\item 64 GB RAM
	\item 500 GB SSD
	\item Intel X550T 10 GbE NIC
	\item Debian 11 Bullseye
\end{itemize}

To distribute the workload generation over the testbed, we spread the Secondaries across all the available isles. We use the first machines of isles A-S and the machines of isle R for workload generation, giving us 10 machines for Secondaries in total. We use the remaining machines for blockchain nodes in different configurations. For simplicity, we vary the number of blockchain nodes as a multiple of 5, since we have 5 machines left from isles A-S.
Same as in AWS setup, for the Primary, we use one of the machines which run the Secondary, as the Primary does not use any resources when the workload is applied to the blockchain network.

\subsection{Inter-switch Communication}

With the local testbed, we focus on finer-grained small-scale experiments which look into how the network scales when the number of blockchain nodes is increased and how the network delay affects the performance of the protocols.

As we have two isles connected to a single switch, we have multiple possible configurations with the experiments involving two, four, or six isles. With two isles, they can either be connected to a single switch or be connected to two different switches. With four isles, two switches can be fully utilized, or there can be a partial utilization of three or four switches. With six isles, either three or four switches can be used. All of these configurations may affect the performance of the blockchain network.

In the next experiments, we send a constant workload of native transfer transactions over 2 minutes to the blockchain network.

\begin{figure}
\centering
  \includegraphics[keepaspectratio=true,scale=0.5110016276450562]{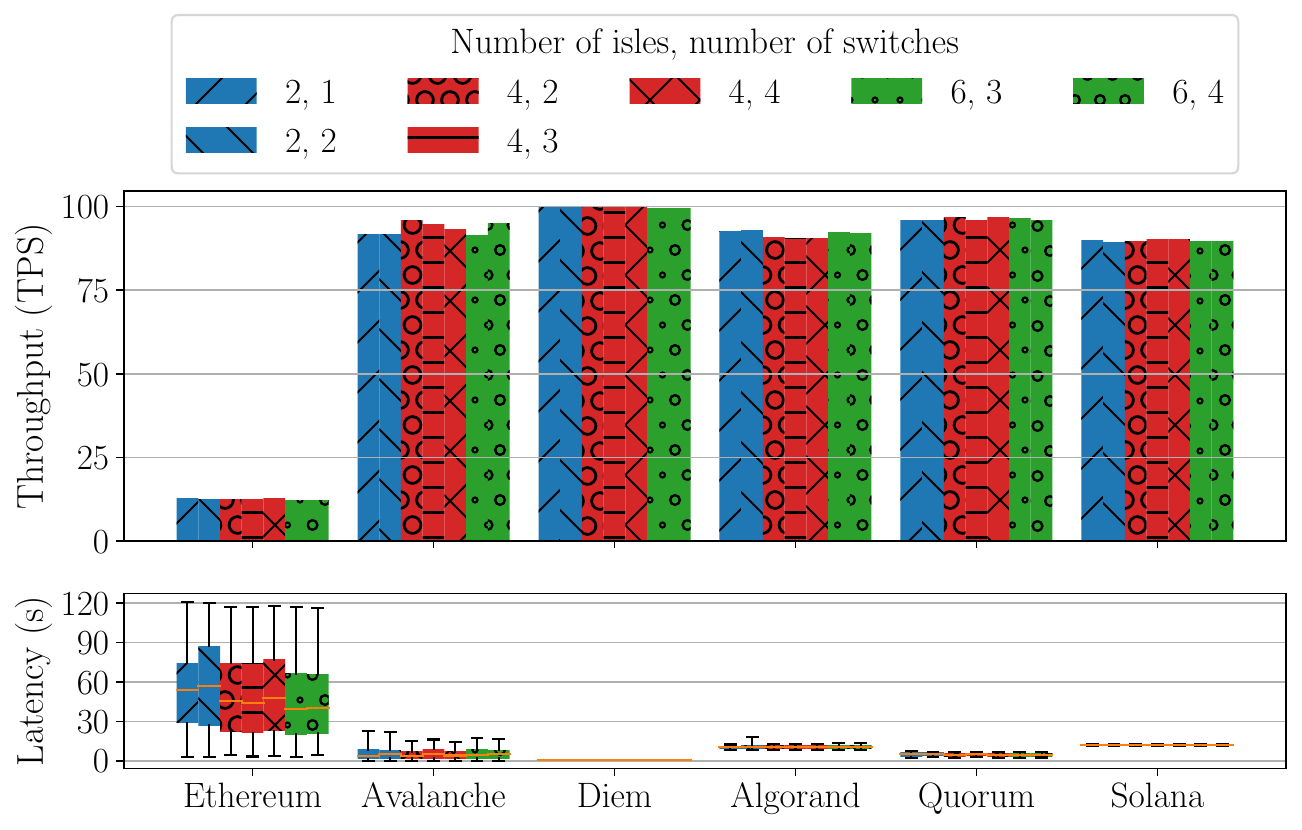}
  \caption{\label{fig:100-switch}Throughput and latency, 100\,TPS workload, varied number of isles and switches}
\end{figure}

Figure~\ref{fig:100-switch} shows the throughput and latency for each blockchain when stressed with a workload of 100\,TPS. We can see that for all of the protocols, the measured throughput stays consistent and is not affected by possible delays added by the switches. For Ethereum, we see a slight decrease in the median latency as we scale up the number of blockchain nodes. We observe that Ethereum achieves significantly lower throughput and displays high latency variance, which can be caused by the default 15 second \texttt{block-period} parameter of Ethereum Clique consensus algorithm.

\begin{figure}
\centering
  \includegraphics[keepaspectratio=true,scale=0.5110016276450562]{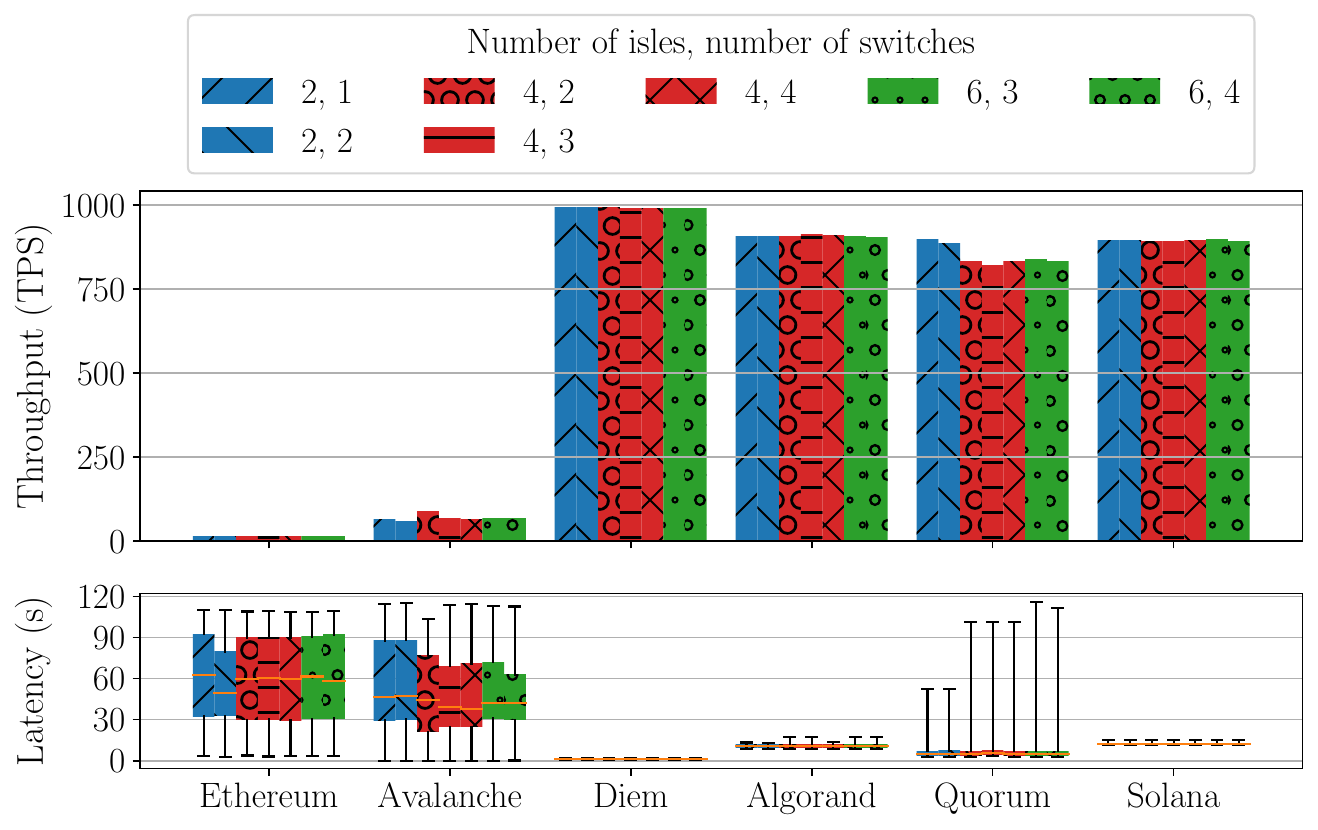}
  \caption{\label{fig:1000-switch}Throughput and latency, 1,000\,TPS workload, varied number of isles and switches}
\end{figure}

Next, we experiment with the same setups and a workload of 1,000\,TPS in Figure~\ref{fig:1000-switch}. Here we start to notice a significant variance in results compared to the previous experiment. First, Avalanche fails to handle the workload, and the latency for the transactions the network manages to commit jumps from 7 seconds to 53 seconds, which is a 7.6 times increase in the average latency. Such behavior could be caused by the default block production configuration, if it is optimized for lower throughput. We also notice that the maximum latency of committed transactions in Quorum starts to depend on the number of nodes. The higher maximum latency means that certain transactions remain in the mempool for longer periods of time, which might be caused by the round-robin block proposer selection.

\begin{figure}
\centering
  \includegraphics[keepaspectratio=true,scale=0.5110016276450562]{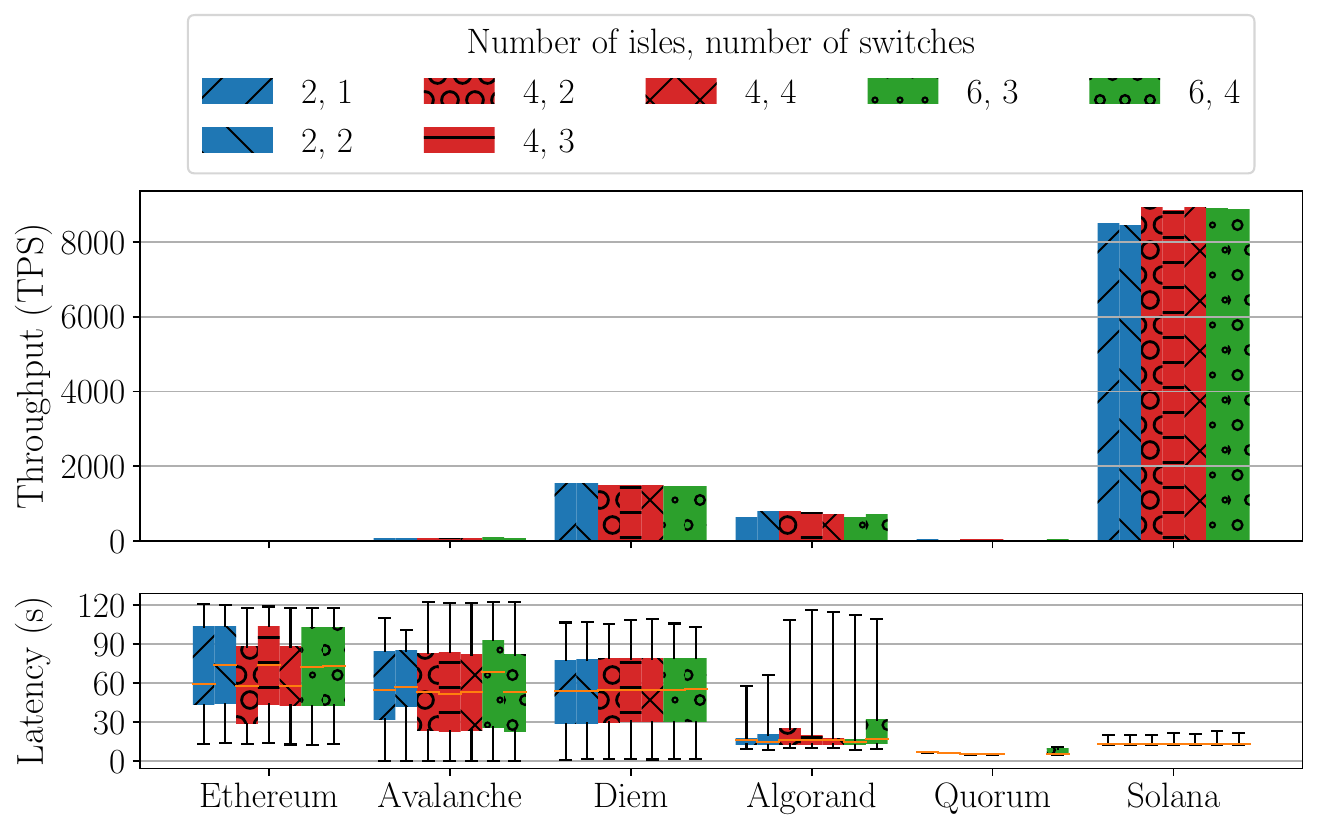}
  \caption{\label{fig:10000-switch}Throughput and latency, 10,000\,TPS workload, varied number of isles and switches}
\end{figure}

Lastly, in Figure~\ref{fig:10000-switch}, we present the results of the experiments with the same setups and a workload of 10,000\,TPS. Here, Solana shows the best results with regard to handling a very high workload. While Quorum exhibits low latency, the throughput is significantly lower than the workload, what might be explained by the internal data structures being oversaturated, and client requests being dropped.

Overall, we make a conclusion that the number of switches does not affect our measurements in a noticeable way and proceed to use the configuration with the minimal number of switches in the next experiments.

\subsection{Isle Scalability}

To evaluate the scalability of the blockchain protocols in terms of the size of the network, we create networks of sizes 5 (1 isle), 10 (2 isles), 20 (4 isles), 30 (6 isles), and 35 (7 isles). We fully utilize the testbed, as it consists of 45 machines in total. We stress the network with the constant workload of native transfer transactions over 2 minutes with a varied rate.

\begin{figure}
\centering
  \includegraphics[keepaspectratio=true,scale=0.5110016276450562]{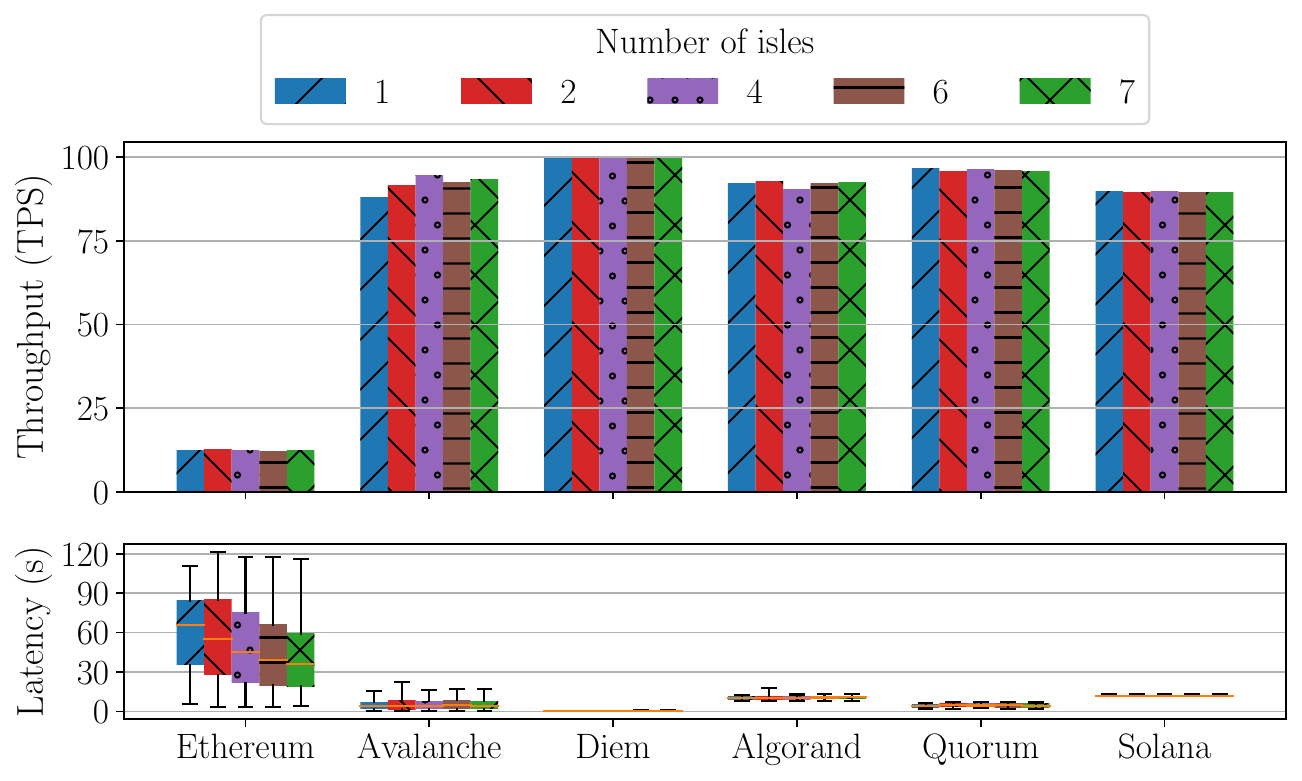}
  \caption{\label{fig:100-scalability}Throughput and latency, 100\,TPS workload, varied number of isles}
\end{figure}

In Figure~\ref{fig:100-scalability}, we compare the latency and the throughput of each protocol under the constant workload of 100\,TPS. We again note that the measured throughput stays consistent between the different sizes of the network for all the blockchains. With Ethereum, we notice a pattern that the median latency tends to become smaller as the network size increases. We cannot increase the size of the network to observe the behavior further. However, the decrease becomes smaller with each network size increase.

\begin{figure}
\centering
  \includegraphics[keepaspectratio=true,scale=0.5110016276450562]{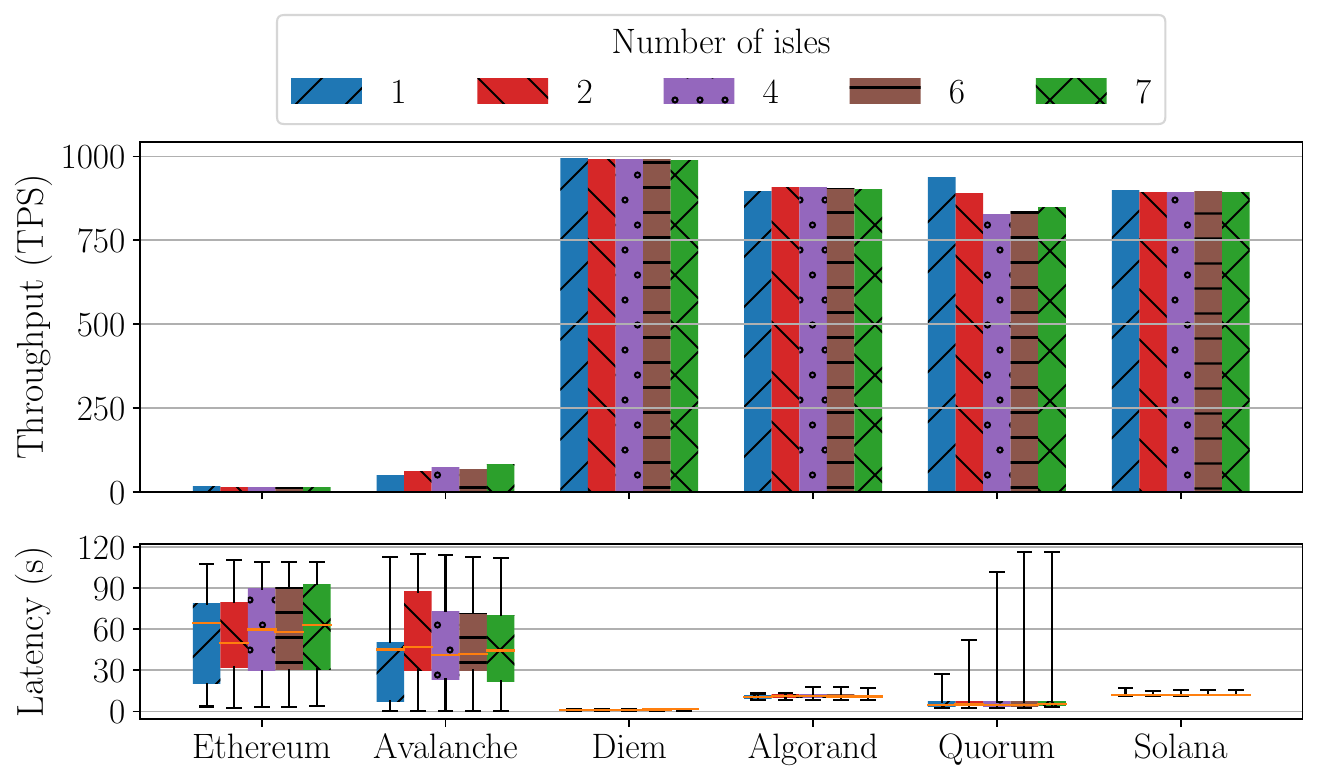}
  \caption{\label{fig:1000-scalability}Throughput and latency, 1,000\,TPS workload, varied number of isles}
\end{figure}

Figure~\ref{fig:1000-scalability} shows the latency and the throughput for the blockchain protocols under test with the 1,000\,TPS workload. The maximum observed latency in Quorum tends to increase with the size of the network. However, at the same time, the throughput does not have a noticeable impact.

\begin{figure}
\centering
  \includegraphics[keepaspectratio=true,scale=0.5110016276450562]{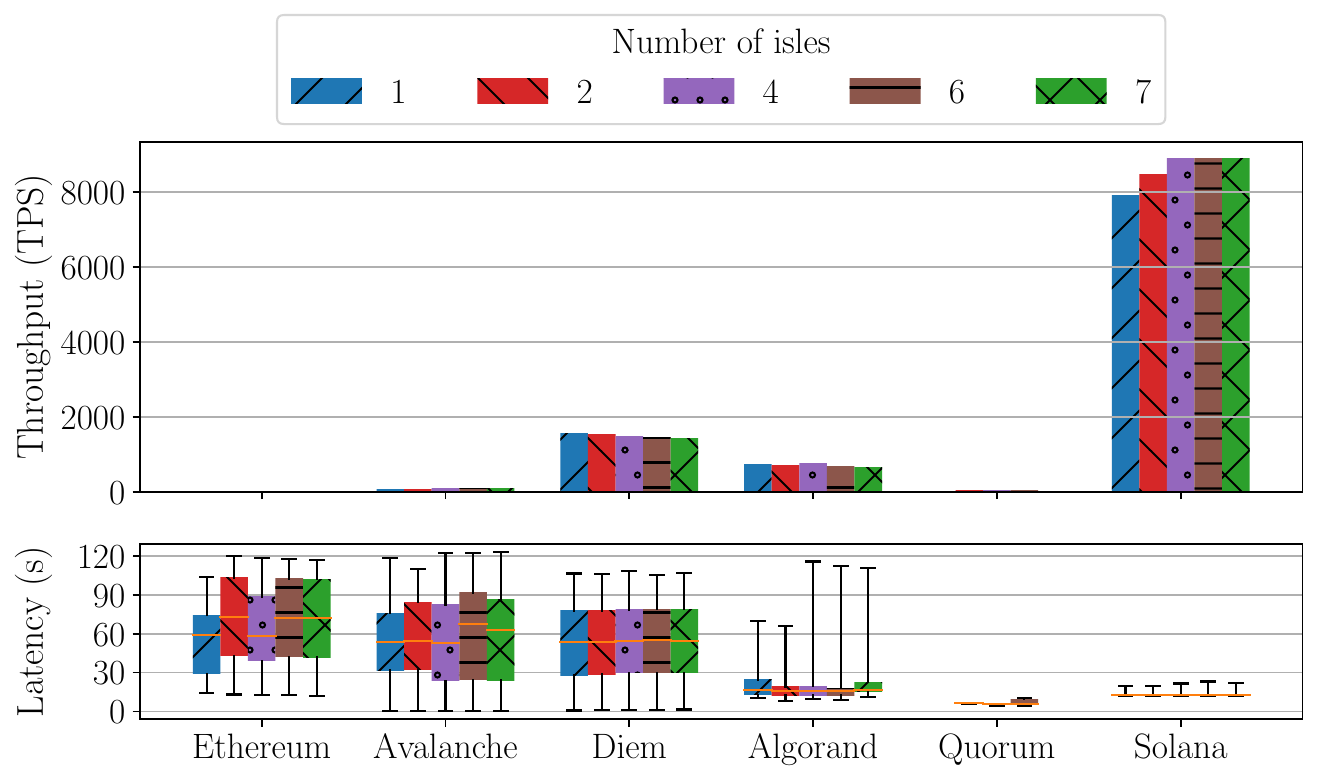}
  \caption{\label{fig:10000-scalability}Throughput and latency, 10,000\,TPS workload, varied number of isles}
\end{figure}

We display the latency and the throughput of all the tested protocols in 7 configurations under a workload of 10,000\,TPS in Figure~\ref{fig:10000-scalability}. We observe an increase in the throughput in Solana as we use 20 nodes. This behavior can be explained by the fact that Solana uses the available processing power of the machines, and its performance scales with the available hardware. For Algorand, we reached its announced peak throughput in the experiment. For Diem, the relatively poor performance compared to Solana can also be explained by the experiment limitation regarding the number of available accounts.

\subsection{Emulated Latency}\label{ssec:emulated}

To evaluate the tolerance against network delays and simulate a real-world geo-distributed environment, we use a network of 35 machines, 7 isles in total. Each of the isles represents a separate location with a fixed delay to other locations. For simplicity, we use equal delay values for all the isles. We experiment with delays of 50, 100, 150, 200, 250, and 300 milliseconds.

\begin{table}
  \centering
  \centering
    \begin{tabular*}{\linewidth}{@{\extracolsep{\fill}} lccccccc}
      \toprule
      Added & 0 & 50 & 100 & 150 & 200 & 250 & 300 \\
      Measured & 1.13814 & 49.758 & 99.663 & 149.689 & 199.73 & 249.782 & 299.727 \\
      \bottomrule
    \end{tabular*}
    \caption{\label{table:measured-delay}Added and average measured RTT (ms) between the isles}
\end{table}

In Table~\ref{table:measured-delay}, we compare the added and measured RTT between the isles in the testbed to verify that the changes we did with \texttt{tc} are correctly applied in the whole network. We see the measured values slightly below the target value because we subtracted the baseline RTT from the added value as we were making the changes.

\begin{figure}
\centering
  \includegraphics[keepaspectratio=true,scale=0.5110016276450562]{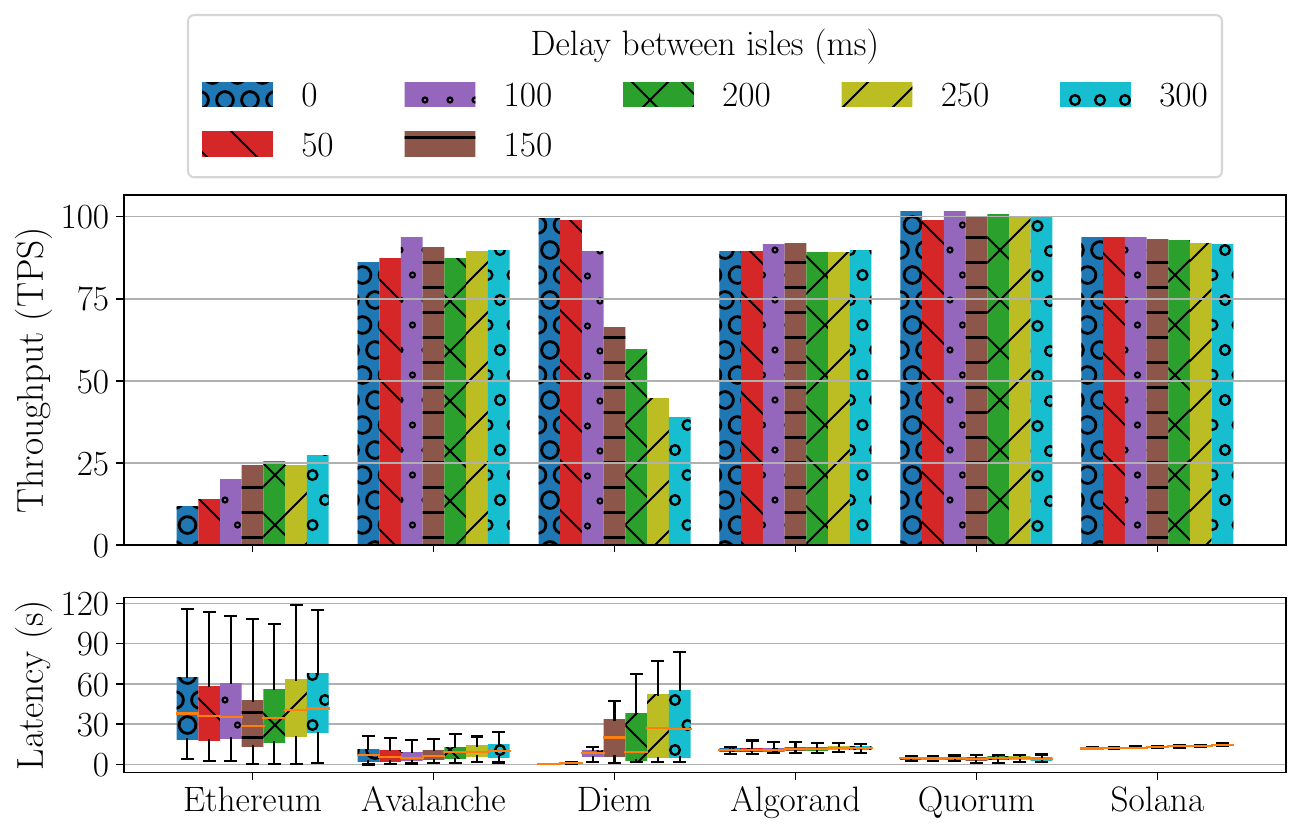}
  \caption{\label{fig:100-delay}Throughput and latency, 100\,TPS workload, varied delay between the isles}
\end{figure}

In Figure~\ref{fig:100-delay}, we compare the throughput and the latency of the protocols under test with 100\,TPS workload, and varied added delay between the isles. We can notice that the performance of Algorand and Quorum stays consistent regardless of the added delay. For Solana, the median latency increases from 12.01 to 14.53 milliseconds. The important observation is that the Diem performance drops significantly with the added delay, and the throughput decreases by more than 50\%. We can infer from the observation that the protocol was optimized for low-latency setups and is not suitable for real-world networks in the current state.

\begin{figure}
\centering
  \includegraphics[keepaspectratio=true,scale=0.5110016276450562]{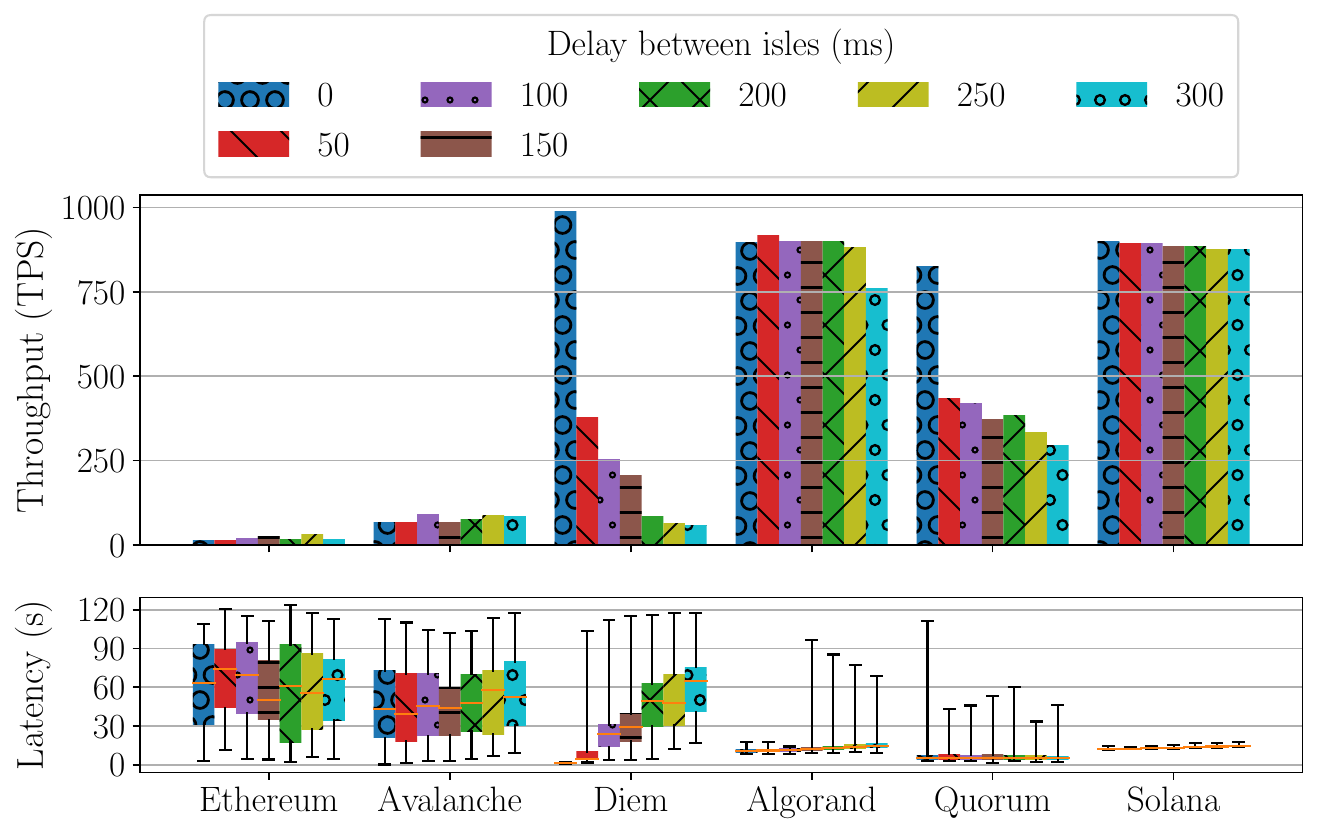}
  \caption{\label{fig:1000-delay}Throughput and latency, 1,000\,TPS workload, varied delay between the isles}
\end{figure}

We display the latency and the throughput for all the tested protocols with 1,000\,TPS workload in Figure~\ref{fig:1000-delay}. Compared to the previous workload, we see that the throughput of Quorum is halved as we add even 50 millisecond delay between the isles. At the same time, the latency stays the same for the different delay settings. For Diem, we see the same behavior of decreased throughput and increased latency. For the other protocols, the performance stays consistent with the increase of the delay in the network.

\begin{figure}
\centering
  \includegraphics[keepaspectratio=true,scale=0.5110016276450562]{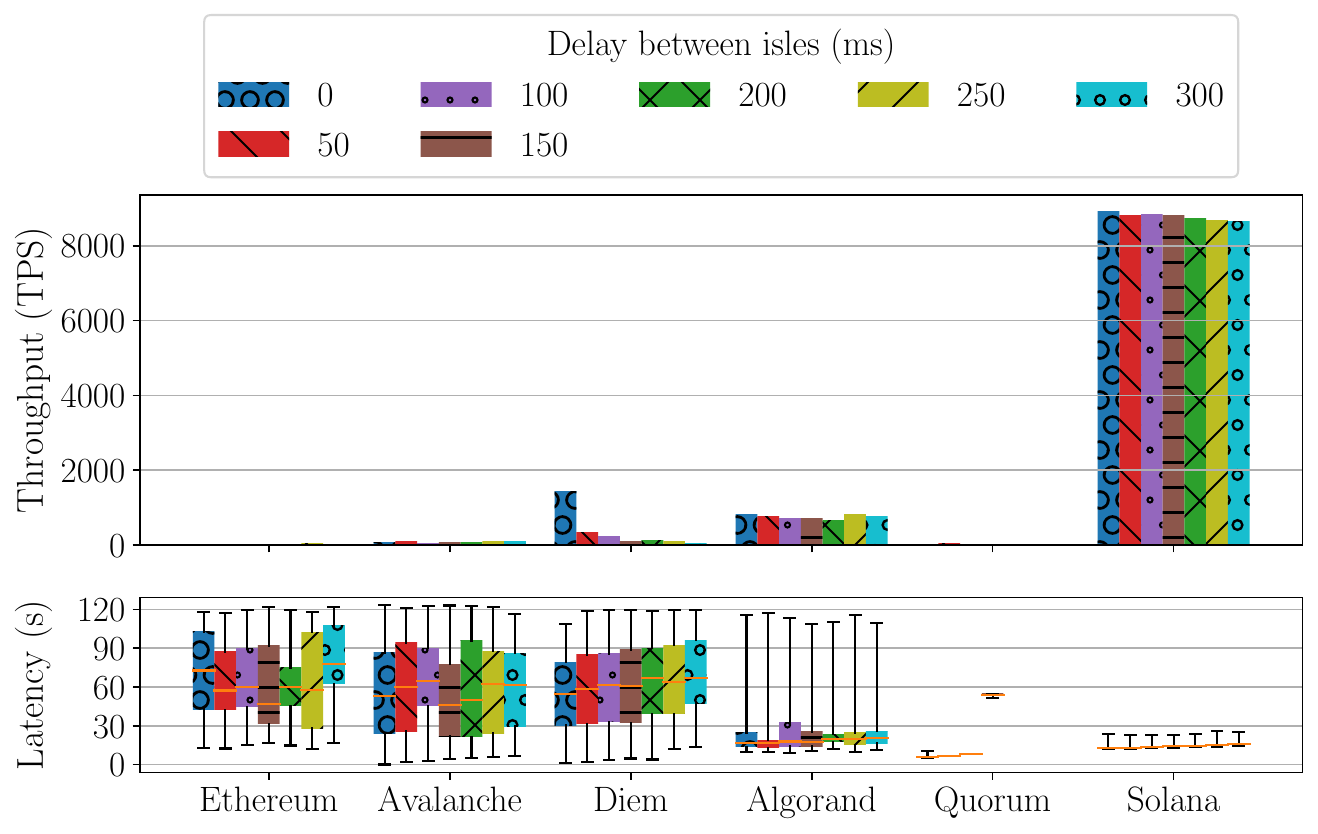}
  \caption{\label{fig:10000-delay}Throughput and latency, 10,000\,TPS workload, varied delay between the isles}
\end{figure}

Figure~\ref{fig:10000-delay} shows the throughput and the latency measures for the protocols under the 10,000\,TPS workload. As before, Ethereum and Avalanche show minimal throughput, and Quorum fails to handle the provided workload. We notice the same performance drop for Diem. Solana and Algorand show consistent performance regardless of the added delay.

For the comparison of the results of the evaluation in AWS and iLab environments, we use the results presented in~\cite{gramoli_diablo_2023}. The corresponding results are displayed in Figure \ref{fig:1000-delay}. As we see in both figures, for the low-latency setups, as \texttt{datacenter} or \texttt{testnet}, Diem reaches its maximum throughput given the provided workload. On the other hand, when we increase the latency between the nodes and use geographically distributed regions as in \texttt{devnet} or \texttt{community}, we observe a significant drop in the throughput of Diem. The results correspond to the experiments with the increased latency in the iLab environment. As for the other blockchains, such as Algorand or Solana, they are optimized for the public networks and increased latencies, and therefore we don't see the drop in throughput, as shown in the figures for AWS and iLab environments.

\section{Discussion}\label{sec:eval-limit}

In this section, we look at the different aspects of the evaluation which can be taken into account in order to increase the depth of understanding of the blockchain protocols. While we performed an extensive set of tests in various environments, there are still more factors and variables that can be changed and which can affect the performance of the protocols.

The work was focused on the behavior of the blockchain protocols under different network conditions, the experiments were limited to the network sizes up to 35 blockchain nodes. Even on smaller scale with emulated delays, we were able to capture similar performance trends to the geo-distributed settings.

We performed the experiments using the default system configurations supplied by the image provider. However, blockchain protocols like Solana recommend\footnote{\url{https://docs.solana.com/running-validator/validator-start\#linux}.} different operating system tuning, such as increasing the size of UDP buffers, as Solana uses UDP for communication, or increasing the limit of memory mapped files. Such tweaks can significantly improve the performance of the protocol but should be examined separately for each protocol.

Protocols such as Algorand have different node types which have different modes of operation. Algorand separates relay nodes and participation nodes, where relay nodes are responsible for communication in the network, and participation nodes participate in consensus. In our deployment scenario, we ran both a relay and a participation node on each machine. Such topology can be suboptimal and not exactly represent a typical deployment. Instead, the Algorand main network can be analyzed, and such topology can be replicated in a private deployment for the performance evaluation.

In our measurements, we calculate the throughput based on the transactions sent by Diablo. We store the hashes of the transactions and compare them to the hashes received in the block subscription or the query for the individual transaction. If the hashes match, we store the commit time of the specific transactions. Such metric only accounts for the transactions generated by Diablo. However, the Solana protocol includes voting transactions into the blocks, meaning that the calculated throughput can be higher if those transactions are included.

Another important point is that while we used the dynamic fee interface for Avalanche, we still observed that some transactions were dropped due to the insufficient fees specified in the transactions. There are multiple possible approaches to solve this issue. On the one hand, we tried to calculate the transaction fees online during the experiment run using the data provided by the blockchain network. It is possible that the approach was not perfect, and therefore the calculation logic can be reviewed and improved. On the other hand, it might be possible to specify static fees in the Avalanche configuration so that they do not provide overhead for the experiment, allowing only to benchmark the raw transaction processing performance. Also, while we experimented with C-Chain, it is important to measure the performance of X-Chain as well.

The differences in the cloud environment and the lab testbed do not allow strict comparison of the metrics. Due to the hardware differences, we can only look at the tendencies and the order of change, but not the exact numbers.

\section{Related Work}\label{sec:related-work}

Dinh et al.~\cite{DWC17} showcase Blockbench, a benchmarking framework designed with the focus on permissioned blockchains. They use a commodity cluster of 48 machines interconnected with a gigabit switch. In the experiments, the number of blockchain nodes is varied from 1 to 32.

Saingre et al.~\cite{saingre_bctmark_2020} use two clusters with Raspberry Pi machines and an HPC-grade cluster. Two network sizes are mentioned without the deployment specification regarding virtualization, containerization, and blockchain node distribution across the physical machines. The used versions of blockchain protocols are not specified as well.

Chacko et al.~\cite{chacko_why_2021} introduce HyperLedgerLab, a benchmarking framework for Hyperledger Fabric. The authors use a Kubernetes cluster consisting of several task-specific nodes. Two cluster setups used for experiments consist of 3 worker nodes which run the Fabric components, and 32 worker nodes respectively. In one of the experiments, the authors introduce an additional network delay of 90 to 110 milliseconds to emulate a geographically distributed environment, and note a negative impact on the performance of the Fabric network.

Nasrulin et al.~\cite{nasrulin_gromit_2022} propose Gromit. They used a cluster of four servers in a single datacenter. The blockchain network sizes from 4 to 128 are mentioned without specifying whether any virtualization or containerization was used, and without mentioning the distribution of blockchain nodes across the four servers. Emulation of geo-distributed setting is present in one of the experiments without specifying the exact configuration of the network emulator. Considering that multiple blockchain nodes are deployed on a single machine as different processes, only filtering by network destination port is possible. This fact leads to the lack of clarity regarding the replication of network delays between the cities, as pings between different pairs of nodes might be different.

In~\cite{ren_bbsf_2023}, Ren et al. run the experiments on a single machine, with three network sizes of 4, 8, and 16 nodes. The authors point out the limitation of creating a large-scale blockchain network on a single machine. It is also reported that Quorum failed to commit transactions with empty smart contract calls given a workload of 500 TPS and a network size of 8 and 16.

Chervinski et al.~\cite{chervinski_analyzing_2023} run their experiments on a network of 5 commodity machines. The authors emulate the ping time of 200 ms between the machines. The number of validators for each of two blockchain protocols varies from 5 to 128, distributed across the machines.

Chacko et al.~\cite{CMF23} outline the need for benchmarking each blockchain in the most suited setup specified in the documentation of this blockchain. This often contradicts with comparing different blockchains on the same ground~\cite{gramoli_diablo_2023}: instead of fine-tuning specific setup for each blockchain one must typically choose the same setup, as realistic as possible, for all blockchains.

\sloppy{Some evaluations compared blockchains on ad hoc benchmarks.
Han et al.~\cite{HSGX20} focused on comparing Ripple, Tendermint, Corda and Hyperledger Fabric to evaluate their scalability potential in the context of Internet of Things. They used the Emulab environment and configured network resources with \texttt{ns-3}.
Shapiro et al.~\cite{SNG20} compared the performance of blockchains relying exclusively on Byzantine fault tolerant consensus protocols, namely Burrow, Quorum and Redbelly Blockchain~\cite{THG23}, on AWS.}

While local clusters and emulated network delays were used in prior work, we use Diablo~\cite{gramoli_diablo_2023} to execute the same set of experiments as in the geo-distributed environment, and allow the comparison of the performance trends of the six state-of-the-art blockchain protocols. Furthermore, we show that tail latency provides additional insight to the impact of underlying protocol implementation and details, such as consensus algorithm.

\section{Conclusion}\label{sec:conclusion}

Our study explored the benefits and drawbacks of cloud environments and local bare metal clusters. Our findings indicated that switches within the LAN had minimal impact on blockchain performance. Additionally, we demonstrated that the average transaction latency of blockchains often failed to accurately represent their tail latency. Notably, the performance patterns observed on our cluster, which incorporated artificial network delays, showed the same behavior as those obtained in geo-distributed settings.

As future work, we plan to evaluate the Redbelly Blockchain~\cite{THG23} on bare metal as it has already been integrated to Diablo~\cite{gramoli_diablo_2023} and showed superior performance than the blockchains we evaluated here.

\subsection*{Acknowledgements}
The authors wish to thank Georg Carle, Holger Kinkelin, Filip Rezabek for their feedback on ealier versions of this paper.
This work is supported in part by the Australian Research Council Future Fellowship funding scheme (\#180100496).

%
%
%
\bibliographystyle{splncs04}
\bibliography{mybibliography}
\end{document}